\documentstyle[12pt]{article}
\begin{document}
\renewcommand{\thepage}{ }
\begin{titlepage}
\title{
\hfill
\vspace{1.5cm}
{\center Magnetization Distribution on Fractals and Percolation
Lattices}
}
\author{
R. M\'elin\\
{}\\
{CRTBT-CNRS, 38042 Grenoble BP 166X c\'edex France}\\
{}\\
{e-mail: melin@crtbt.polycnrs-gre.fr}}
\date{}
\maketitle
\begin{abstract}
\normalsize
We study the magnetization distribution of the Ising model
on two regular fractals (a hierarchical lattice, the regular simplex)
and percolation clusters at the percolation threshold in a two
dimensional imbedding space.
In all these cases, the only fixed point is $T=0$. In the case
of the two regular fractals, we show that the magnetization distribution
is non trivial below $T^{*} \simeq A^{*}/n$, with $n$ the number
of iterations, and $A^{*}$ related to the order of ramification.
The cross-over temperature $T^{*}$ is to be compared with
the glass cross-over temperature
$T_g \simeq A_g/n$. An estimation of the ratio $T^{*}/T_g$ yields
an estimation of the order of ramification
of bidimensional percolation
clusters at the threshold ($C = 2.3 \pm 0.2$).
\end{abstract}
\end{titlepage}

\newpage
\renewcommand{\thepage}{\arabic{page}}
\setcounter{page}{1}
\baselineskip=17pt plus 0.2pt minus 0.1pt
\section{Introduction}
We consider the problem of Ising spin systems on a family of fractals
with a zero temperature fixed point. More especially, we consider
the case of a hierarchical lattice, the regular simplex (regular
fractals), and bidimensional percolation clusters at the percolation
threshold. This problem has its origin in the pioner work of Mandelbrot
\cite{Mandel}. The problem is also of an
experimental interest because of the existence of random magnets
with a fractal structure (see for instance \cite{Rev}).
As a byproduct, we derive a measure of the
ramification order of percolating clusters. To do so, we compare
to cross-over temperatures: 1) the glass cross-over temperature $T_g$
2) the magnetization distribution cross-over temperature $T^{*}$.
The dominant behavior of the glass cross-over temperature was
discovered in the 80's \cite{ref7} \cite{Henley}. This quantity
is related to the correlation length of the magnet. Below $T_g$,
the correlation length is larger than the system size, leading
to a linear regime of the logarithm of the relaxation time
as a function of the inverse temperature. Above $T_g$, the correlation
length is smaller than the linear size of the cluster, leading to
a parabolic behavior for the logarithm of the relaxation time
as a function of the inverse temperature. This behavior is well
established on a numerical basis \cite{ref8}.
The second cross-over temperature $T^{*}$
was not much studied in the past.
Below $T^{*}$ [which is inverse proportional to the logarithm of the
number of sites], the magnetization distribution is non Gaussian.
These non Gaussian magnetization distributions also accur on the
ferromagnetic Cayley tree \cite{Melin}, however with a different decay
with the system size. More surprising is the appearance of similar
magnetization distributions in magnetized spin glasses \cite{Chayes}.
However, in real spin glasses, this non Gaussian magnetization distribution
is not expected to vanish in the thermodynamic limit as it is the
case on ferromagnetic systems. We first study the case of regular
fractals (a hierarchical lattice and the regular simplex). We use the
Swendsen algorithm to generate equilibrium states \cite{swendsen},
and thus to calculate
the magnetization distributions. In the case of the regular simplex,
it is possible to determine the geometric origin of the different
local maxima of the magnetization distribution at low temperatures.
In section \ref{general}, we make a general argument to
estimate the magnetization distribution cross-over scale, which is
shown to be inverse proportional to the number of sites,
and proportional to the ramification order $C$.
In the case of percolating clusters, we calculate the magnetization
distribution at low temperature. Averaging over the geometry, we can
see the existence of finite size effects: the spin system is more
ferromagnetic if one lowers the number of generations. We also studied
the correlation between different equilibrium states, which gives
us an other way to estimate $T^{*}$.
Finally, we calculate the ration $T_g/T^{*}$, which yields
an estimation of the ramification order of bidimensional percolation
clusters. We end up with some concluding remarks.

\section{A Hierarchical Lattice}
\label{hier}
The interest of hierarchical lattices is that the thermodynamics
of the Ising model is exactly soluble on these structures.
We consider the case of the hierarchical lattice built with the
rules of figure \ref{fig1}. The coordination of every site is
bounded above, which is not always the case for hierarchical lattices
(for instance for the hierarchical diamond, in which case the
coordination diverges in the thermodynamic limit).
The number of sites of the hierarchical lattice of figure \ref{fig1}
is $N_n = 4. 6^{n} /5 + 6/5$, whereas the
end to end distance is $L_n = 4^{n-1}$, leading to the
fractal dimension $\overline{d} = \ln{6}/\ln{4}$.

An interesting aspect of hierarchical lattices is that their
Yang and Lee zero can be calculated easily in the temperature plane.
This was done in \cite{ref1} in the case of the hierarchical diamond.
The Yang and Lee zeros set is a Julia set. For comparison with
the results of \cite{ref1}, we also calculate the Yang and Lee zeros of
our hierarchical lattice.
First, we need to derive recursion relations for the partition function.

\subsection{Recursion for the Partition Function}
The technique (standard for hierarchical lattices) consists in
carrying out the trace over the the spin variables at the smallest
scales, and to derive the renormalization of the temperature and magnetic
field. The partition function of the hierarchical lattice with one generation
is
\begin{equation}
Z_1(\Sigma,\Sigma') = \sum_{\sigma_0,\sigma_1,\sigma_2,\sigma_3}
e^{\beta J(\sigma_0 \Sigma + (\sigma_0+\sigma_3)(\sigma_1+\sigma_2))
+ \sigma_3 \Sigma'}
e^{\beta H(\sigma_0+...+\sigma_3)}
,
\end{equation}
where the spins are pictured on figure \ref{fig1}.
A straightforward calculation leads to
\begin{eqnarray}
Z_1(\Sigma,\Sigma') &=& 2 \left( e^{4 \beta J} \cosh{(\beta J(\Sigma+\Sigma')
+ 4 \beta H)}
+  e^{-4 \beta J} \cosh{\beta J(\Sigma+\Sigma')} \right.
\nonumber\\
&&
\label{eq1}
\left.
 + \cosh{(\beta J(\Sigma-\Sigma')+2 \beta H)}
+ \cosh{(\beta J(\Sigma-\Sigma')-2 \beta H)} \right.\\
\nonumber
&& \left. +
2 \cosh{\beta J(\Sigma-\Sigma')}
+ 2 \cosh{(\beta J(\Sigma+\Sigma')+2 \beta H)} \right)
.
\end{eqnarray}
The partition function $Z_1(\Sigma,\Sigma')$ can be rewritten under
the form
\begin{equation}
Z_1(\Sigma,\Sigma') = {\cal N} e^{\beta \tilde{J} \Sigma \Sigma'}
e^{\beta \tilde{H}(\Sigma+\Sigma')}
\label{eq2}
.
\end{equation}
We have introduced here three parameters ${\cal N}$,
$\tilde{J}$ and $\tilde{H}$
for three distinct partition functions $Z(+,+)$, $Z(-,-)$ and
$Z(+,-)=Z(-,+)$. The partition function can thus be consistently
be brought under the form (\ref{eq2}).
The identification of (\ref{eq1}) and (\ref{eq2}) leads to
\begin{eqnarray}
\label{eq5}
{\cal N} &=& \left( Z_1(+,+) Z_1(+,-)^{2} Z_1(-,-) \right)^{1/4}\\
\label{eq6}
\exp{\beta \tilde{J}} &=& \frac{{\cal N}}{Z_1(+,-)}
= \frac{\left( Z_1(+,+) Z_1(-,-) \right)^{1/2}}{{\cal N}}\\
\label{eq7}
\exp{4 \beta \tilde{H}} &=& \frac{Z_1(+,+)}{Z_1(-,-)}
.
\end{eqnarray}
In the case of a zero magnetic field, we obtain
\begin{equation}
Z_n(J) = {\cal N}^{6^{n-1}} Z_{n-1}(\tilde{J})
,
\end{equation}
with
\begin{equation}
{\cal N}^{2} = 8 \left( \cosh{6 \beta J} + 3 \cosh{2 \beta J} + 4 \right)
\left( \cosh{4 \beta J} + 2 \cosh{2 \beta J} +1 \right)
,
\end{equation}
and
\begin{equation}
e^{2 \beta \tilde{J}} = \frac{ \cosh{6 \beta J} +
3 \cosh{2 \beta J} + 4}
{2\left( \cosh{4 \beta J} + 2 \cosh{2 \beta J} + 1 \right)}
\label{eq3}
.
\end{equation}
The renormalization equation (\ref{eq3}) has only one fixed
point: $J=0$, so that the spin system is always disordered at any
finite temperature in the thermodynamic limit.

\subsection{Digression: Yang and Lee zeros in the Temperature Plane}
We now calculate the Yang and Lee zeros in the temperature plane.
This is not the main subject of this paper, but it is interesting to
compare with existing results on the hierarchical diamond, where the
Yang and Lee zero in the temperature plane form a Julia set \cite{ref1}.
We work in the plane of the variable $z=\exp{2 \beta J}$ and invert the
relation (\ref{eq3}), which leads to a polynomial of degree $6$:
\begin{equation}
z^{6} - 2 \tilde{z} z^{5} + (3-4 \tilde{z}) z^{4}
+ 4(2-\tilde{z}) z^{3} + (3-4 \tilde{z}) z^{2} - 2 \tilde{z} x
+ 1 = 0
\label{eq4}
,
\end{equation}
If $z$ is a solution of (\ref{eq4}), $1/z$ is a solution too, so
that the set of zeros of the partition function is invariant
under the inversion $z \rightarrow 1/z$. In order to compute the set
of zeros of the partition function, we use the methods of \cite{ref1},
that is we start from a given point in the complex plane,
we calculate the zeros of (\ref{eq3}), choose one solution among the
six zeros at random, and reiterate.
The zeros are computed using a Laguerre's method \cite{ref2}.
The resulting set is independent on the initial value
of $z$, as far as one eliminates the first zeros.
The result is plotted on figure \ref{fig2}. The set of zeros are
concentrated on lines, which intersect the real axis in the vicinity of
the point $z=0$. A zoom in the vicinity of $z=0$ reveals that
the density of zeros vanishes as one approaches the origin (see
figure \ref{fig3}). Contrary to the case of the hierarchical diamond,
the Lee and Yang zeros are concentrated on lines in the
$z=\exp{2 \beta J}$ plane, whereas the Julia set of the hierarchical
diamond is a fractal.

\subsection{Glass Cross-over Temperature}
The glass temperature cross-over $T_g$ is related to the decay
of the correlation function from one end of the fractal to the other
\cite{ref7} \cite{ref8} \cite{ref9}. Since the Ising model is soluble
on the hierarchical lattice, it is easy to get the behavior of the
correlation length between the extremal sites as a function of
the number $n$ of iterations and the temperature.
We call $x$ the dimensionless inverse temperature $x=\beta J$.
In the low temperature regime, we get from equation (\ref{eq3})
$x_{n+1} \simeq x_n - \ln{2}/2$,
where $x_n$ is the value of $x$ after $n$ renormalization steps.
In the low temperature regime, $x_{n+1} \simeq 2 x_n^{4}$.
The limit between the high and low temperature regimes corresponds
to the minimum $\kappa$ of $2 x^{4} - x + \ln{2}/2$. We find
$\kappa \simeq 0.59$. The correlation between the two extremal
sites is nothing but
\begin{equation}
\langle \Sigma \Sigma' \rangle
= \frac{1}{Z[0]} \frac{\partial}{\partial \mu}
Z[\mu = 0]
,
\end{equation}
where 
\begin{equation}
Z[\mu] \propto \sum_{\Sigma,\Sigma'} e^{\beta J_n \Sigma \Sigma'}
e^{\mu \Sigma \Sigma'}
,
\end{equation}
so that the correlation is simply $\langle \Sigma \Sigma'
\rangle = \tanh{\beta J_n}$, which decreases from unity to zero
as $n$ increases from 1 to $+ \infty$.
The cross-over temperature $T_g$
corresponds to $x_{n}> \kappa$ and $x_{n+1}< \kappa$,
leading to the dominant behavior of the glass cross-over temperature
\begin{equation}
T_g = \frac{2 J}{n \ln{2}}
.
\end{equation}

\subsection{Magnetization Distribution}
The magnetization distribution becomes non gaussian below a
temperature $T^{*}$. This is due to the existence of macroscopic
domains that are weakly connected ot the rest of the structure.
The temperature cross-over $T^{*}$ is evaluated in section
\ref{general} in a more general context, so that we do not reproduce
the argument here in the case of the hierarchical lattice.
The result in the case of the hierarchical lattice is
\begin{equation}
T^{*} = \frac{4 J}{n \ln{6}}
\label{T*1}
,
\end{equation}
where we have kept only the dominant behavior for large $n$.
The results for the magnetization distribution are plotted on figure
\ref{fig4} for $n=4$. Using (\ref{T*1}) for $n=4$, we get
$T^{*} \simeq 0.55$, which is consistent with the data of figure
\ref{fig4}. Notice that, in this case, $T_g<T^{*}$.
\section{Regular Simplex}
\label{simplex}
The regular simplex recursion is plotted on figure \ref{fig5}.
The number of sites if $N_n=3^{n}$, and the corresponding distance is
$L_n=2^{n}$, so that the fractal dimension is $\overline{d}=\ln{3}/
\ln{2}$.
\subsection{Recursion for the Partition Function}
The partition function in a zero external field
can be calculated recursively by calculating
the trace over the spins at the deepest generation.
With the notations of figure \ref{fig5}, the partition function
is
\begin{eqnarray}
Z(\Sigma_1,\Sigma_2,\Sigma_3) &=&
\exp{\left(\beta J(\Sigma_1(\sigma_{12}+\sigma_{13})+\Sigma_2(\sigma_{21}+
\sigma_{23}+\Sigma_3(\sigma_{31}+\sigma_{32})\right)}\\
&& \exp{\left(\beta J(\sigma_{31} \sigma_{32}+ \sigma_{21} \sigma_{23}+
\sigma_{12} \sigma_{13})\right)}
.
\end{eqnarray}
We look for $Z$ under the form
\begin{equation}
Z(\Sigma_1,\Sigma_2,\Sigma_3) = {\cal N} \exp{\left( \beta \tilde{J}
(\Sigma_1 \Sigma_2 + \Sigma_2 \Sigma_3 + \Sigma_1 \Sigma_3)\right)}
.
\end{equation}
The renormalized form of $Z$ has two parameters ($N$ and $\tilde{J}$),
for two distinct equations: one for $(\Sigma_1,\Sigma_2,\Sigma_3)
\in {\cal S}_1$, and the other for $(\Sigma_1,\Sigma_2,\Sigma_3)
\in {\cal S}_2$, with
\begin{eqnarray}
{\cal S}_1 &=& \{(+,+,+),(-,-,-)\}\\
{\cal S}_2 &=& \{(+,+,-),(+,-,+),(-,+,+),(-,-,+),(-,+,-),(+,-,-)\}.
\end{eqnarray}
The parameters ${\cal N}$ and $\tilde{J}$ are simply given by
\begin{eqnarray}
{\cal N} &=& \left( Z(+,+,+) Z(+,+,-) \right)^{1/4}\\
\beta \tilde{J} &=& \frac{1}{4} \ln{ \left(\frac{Z(+,+,+)}{Z(+,+,-)}
\right)}
.
\end{eqnarray}
Carrying out the trace over the $\sigma$ variables leads to
\begin{eqnarray}
Z(+,+,+) &=& 27 e^{-3 \beta J}+27 e^{\beta J} + 9 e^{5\beta J}
+ e^{9 \beta J}\\
Z(+,+,-) &=& 3 e^{-7 \beta J}+19 e^{-3 \beta J}+33 e^{\beta J}
+ 9 e^{5 \beta J}
.
\end{eqnarray}
In the high temperature regime, the coupling constant renormalization
is $\tilde{x} \simeq x^{3}$ and, in the low temperature regime,
$\tilde{x} \simeq x - (\ln{3})/2$, where $x=\beta J$.
The cross-over between the high and low temperature regime is
$x=\kappa$, with $\kappa$ the minimum of $x^{3}-x+(\ln{3})/2$
($\kappa \simeq 0.58$).
The correlation function is
\begin{equation}
\langle \Sigma_1 \Sigma_2 \rangle = \frac{1}{Z[0]}
\frac{\partial}{\partial \mu} Z[\mu=0]
,
\end{equation}
with
\begin{equation}
Z[\mu] \propto \sum_{\Sigma_1,\Sigma_2,\Sigma_3}
\exp{\left(x_n (\Sigma_1 \Sigma_2+\Sigma_1 \Sigma_3 +\Sigma_2
\Sigma_3 \right)}
\exp{\left(\mu \Sigma_1 \Sigma_2 \right)}
= \frac{e^{3 x_n} - e^{-x_n}}{e^{3 x_n}+3 e^{- x_n}}
,
\end{equation}
from what we deduce the dominant behavior of the cross-over temperature
scale
\begin{equation}
T_g = \frac{2 J}{n \ln{3}}
.
\end{equation}
\subsection{Magnetization Distribution}
The magnetization distribution is plotted on figure \ref{fig6}
for various temperatures. Again, we do not reproduce the calculation
of the magnetization cross-over temperature $T^{*}$ since a general argument
in developed in section \ref{general}. The result for the dominant
behavior of $T^{*}$ in the limit of large $n$ is
\begin{equation}
T^{*} = \frac{6 J}{n \ln{3}}
\label{T*2}
.
\end{equation}
In the case $n=7$ (which corresponds to figure \ref{fig6}), we have
$T^{*} \simeq 0.8$, which is consistent with the data of figure \ref{fig6}.
Notice also that, in the low temperature regime, when the non gaussian
structure of the magnetization distribution is fully developped,
it is possible to identify the origin of the different local maxima
of $P(m)$.
To do so, we carry out a low temperature expansion of the magnetization
distribution around the broken symmetry state $m=1$,
up to the order $\mu^{3}$, where
$\mu = e^{-\beta J}/(e^{\beta J}+e^{-\beta J}) \simeq e^{-2 \beta J}$.
The result is
\begin{eqnarray}
\label{Pm}
P(m) &=& (1-n \mu^{2} - A \mu^{3}) \delta (1-m) + 3 \mu^{2}
\sum_{p=0}^{n-1} \delta(m-1+2. 3^{p-n})\\
&& + \mu^{3} \left[ \sum_{p=0}^{n-2} (3^{n-p}-3) \delta(m-1+2.3^{p-n})
+ 6 \sum_{M\in {\cal A}} \delta(1-\frac{2 M}{3^{n}}) \right]
+ o(\mu^{3})
.
\end{eqnarray}
$A$ is a normalization constant and the set ${\cal A}$ is
\begin{equation}
{\cal A} = \{M\in \langle 1,..., [N/2] \rangle, \exists k,
\exists l, 0 \le l <k, M=3^{k} \pm 3^{l} \}
.
\end{equation}
We can recognize on figure \ref{fig6} the contributions from
$m=1/3,1/9,1/27$ but also from $m=1/3 \pm 1/27, 1/3 \pm 1/9, 1/9 \pm 1/27$.

\section{General Argument for Regular Fractals with a Zero Temperature 
Transition}
\label{general}
We first recall what happens on the Cayley tree \cite{Melin}, where there
exists also a glassy cross-over, with $T_g \propto J/\ln{n}$, where $n$
is the number of generations of the tree, and also a crossover in the
magnetization distribution, with $T^{*} \propto J/\ln{n}$. It is
possible to derive exact recursion relations for the average magnetization,
to solve the recurence, and to obtain the temperature cross-over $T^{*}$.
The cross-over temperature is interpreted as follows: below $T^{*}$,
there exists less than one broken bound on a path connecting the center of
the tree to the leaves, which leads to the criterium for the magnetization
distribution cross-over temperature:
\begin{equation}
n \mu \simeq 1.
\label{tree}
\end{equation}
Taking the logarithm leads to $T^{*} \propto J/\ln{n}$.

In the case of fractals, there are not bifurcations, so that the
criterium corresponding to (\ref{tree}) is
\begin{equation}
\mu^{C} N_n \simeq 1
,
\end{equation}
which means that, at the cross-over, there exists only $C$ broken bounds
on the graph, where $C$ is the number of bounds that need to be cut to isolate
a $p$ generations fractals in the bulk of a $n$ generations fractal. This
is the ramification order.
We thus obtain
\begin{equation}
T^{*} = \frac{2 J C}{\ln{N_n}}
\label{cross}
.
\end{equation}
In order to check this equation, we come back to the case of the hierarchical
lattice, where we obtained recursively the partition function in an
external magnetic field. We impose a small magnetic field, and calculate
the partition function in the presence of the small magnetic field,
and deduce the average magnetization numerically as a function of the
number of generations. For a given number of generation, we look
for the temperature for which the magnetization is a given constant.
We thus obtain the curve $\beta^{*}(n)$, which is expected to be a
straight line from equation (\ref{cross}). This is indeed the case,
as shown on figure \ref{supp}, which validates the previous
heuristic argument.
\section{Percolation Clusters at the Threshold}
\subsection{Glass cross-over temperature}
Percolation clusters at the percolation threshold were studied in
the past as an exemple of critical dynamics, with a glassy-like dynamics
below $T_g$ \cite{ref7} \cite{ref8} \cite{ref9}.
The glass cross-over temperature $T_g$ is evaluated in \cite{ref7}
with the argument that below $T_g$, the correlation length
\cite{coniglio1}
\begin{equation}
\xi_T = \exp{\left( \frac{2 J \nu_p}{T} \right)}
\end{equation}
is inferior to the linear size of the cluster. One gets easily \cite{ref7}
\begin{equation}
T_g = \frac{2 J \overline{d} \nu_p}{\ln{N}}
\label{Tgperco}
,
\end{equation}
where $N$ is the number of sites of the percolation cluster.

\subsection{Magnetization Distribution}
The magnetization distribution of a single cluster at low temperatures
looks like the one of regular fractals, apart from the fact that the
localization of the local minima depends on the geometry and is not
easy to localize. For instance, we generated a percolation
cluster at threshold in a bidimensional imbeding space ($p_c = 0.593$).
The corresponding low temperature magnetization distribution is plotted
on figure \ref{fig7} for various temperatures.
We show on figure \ref{fig8} two configurations
belonging to the local maximum with a normalized number of up spins
$N_{\uparrow}/N \simeq 0.38$.
The interpretation of the magnetization distribution thus depends
crucially on the geometry. It is therefore legitimate to ask whether
the local maxima structure persists after averaging over the geometry.
The answer is `no': we plotted on figure \ref{fig9} the magnetization
distribution averaged over the geometry for different temperature.
We observe no local maxima. If the temperature increases,
the weight of the ferromagnetic maximum decreases, whereas
the tail of the distribution increases, which is consistent with the
fact that the magnetization distribution is gaussian in the high
temperature regime.

Moreover, by averaging over the geometry, we can adress the question
of finite size effects. We plotted on figure \ref{fig10} the distribution
of magnetization for two different sizes at the same temperature.
We see that the small clusters have a more pronounced ferromagnetic peak,
which is in agreement with the fact that $T^{*}$ decreases with the
system size.
\subsection{Correlations}
We wish to analyze the correlation between the different spin
configuration at a given temperature. To do so, we introduce $n$
replica of the spin system. We call $q_i^{(\alpha)}\in\{0,1\}$
the value of the Potts variables in the replica $\alpha$
($1 \le \alpha \le n$), and we consider the following correlation:
\begin{equation}
R_N(q,T)= \frac{1}{q} \sum_{i=1}^{N} \frac{2}{n(n-1)}
\sum_{\langle \alpha,\beta \rangle} q_i^{(\alpha)}
q_i^{(\beta)}
\end{equation}
for a given geometry with a given magnetization (all the replica have the
same magnetization).
The variable q is the average magnetization:
\begin{equation}
\forall \alpha, q=\frac{1}{N} \sum_{i=1}^{N} q_i^{(\alpha)}
.
\end{equation}
The correlation $R_N(q,T)$ in a sector of given magnetization
is nothing but a kind of Edwards-Anderson order parameter
in a subspace of given magnetization. In the high temperature regime,
$q_i^{\alpha}$ is $0$ with a probability $1/2$ and $1$ with a probability
$1/2$, so that the average of $q_i^{(\alpha)} q_j^{(\alpha)}$ is $1/4$,
so that
\begin{equation}
R_N(q,t)  = \frac{N}{4 q}
\end{equation}
if $T>T^{*}$.
We are interested in the variations of $R$ for a given magnetization as a
function of the temperature. These variations are plotted on figure
\ref{fig11} for a given geometry. The magnetization is
$q=N_{\uparrow}/N \simeq 0.38$. The four different curves are plotted
for neighboring magnetizations. As expected, $R$ tends to $0.38$
above $T^{*}$. The estimation of $T^{*}$ from this technique is
coherent with the estimation from the analysis of the magnetization
distribution (see figure \ref{fig7}).

\subsection{Evaluation of the Ramification Order of Percolating Clusters}
We know from two different methods that the cross-over temperature
$T^{*}$ for $N=2660$ sites is about $T^{*} \simeq 0.6$. From this data,
we can deduce the ramification order $C$, which, in the case of regular
fractals was interpreted as the number of links to be cut
to isolate a $p$ generations fractal inside a $n$ generations fractal
($n>p$). We deduce from (\ref{cross}) and (\ref{Tgperco}) that
\begin{equation}
\frac{T^{*}}{T_g} = \frac{C}{\nu_p \overline{d}}
\label{eqx}
.
\end{equation}
On the other hand, we know that $\nu_p \simeq 1.3$ in a bidimensional
imbedding space \cite{Stauffer}, and that $\overline{d} = 1.78 \pm 0.02$
\cite{Stanley}.
With $N=2660$ sites, we evaluate $T_g \simeq 0.6$ from (\ref{Tgperco}).
Using (\ref{eqx}), we get $C=2.3 \pm 0.2$. The incertainty comes from the
incertainty in the location of the cross-over temperature $T^{*}$.
This result is in agreement with
the fact that the order of ramification of percolating clusters
is believed to be finite and superior to two (percolating clusters
are not quasi one dimensional) \cite{Kirk}.
\section{Conclusion}
We have studied the magnetization distribution on fractals with a zero
temperature transition. Below a temperature inverse proportional to
the logarithm of the number of sites, and proportional to
the order of ramification, the magnetization is not trivial,
with local maxima. This is due to the fact that is is possible to cut the
graph into many large parts, by breaking only a small number of links.
We have introduced a magnetization-dependent Edwards-Anderson order
parameter, the variations of which are correlated with the
magnetization distribution. It would be interesting to have an
analytical proof of this fact.
We have shown in that the cross-over temperature
$T^{*}$ of percolating clusters is related to the order of ramification
of the structure. We find $C=2.3 \pm 0.2$ for bidimensional percolation
clusters. The cross-over temperature $T^{*}$ can be estimated in two
different ways: 1) by calculating directly the probability distribution
at different temperatures 2) by calculating the correlation between
equilibrium states with a given magnetization. Our result is
in agreement with the fact that the percolation cluster at threshold
is not quasi one dimensional (which was known a long time ago !).
However, it is clear that our method to determine the order of ramification
is not accurate, since one has to estimate a cross-over temperature,
which is determined up to a certain incertainty.
It would be interesting to generalize this study to
magnetized spin glass phases. This will be done in a near future.

I acknowledge J.C. Angl\`es d'Auriac who lent me his Swendsen program,
and I thank B. Dou\c{c}ot for comments on the manuscript.

\newpage
\renewcommand\textfraction{0}
\renewcommand
\floatpagefraction{0}
\noindent {\bf Figure captions}

\begin{figure}[h]
\caption{}
\label{fig1}
Construction of the hierarchical lattice. At the $n=0$ step, we
start with two sites connected. After one step, the lattice contains 6
sites. At each step, the links are transforms as in the
$n=0 \rightarrow n=1$ transformation.
\end{figure}

\begin{figure}[h]
\caption{}
\label{fig2}
Set of zeros of the partition function in the plane $z=\exp{2 \beta J}$
We have plotted 39000 zeros calculated by the procedure described
in the text. 1000 zeros have been eliminated.
\end{figure}

\begin{figure}[h]
\caption{}
\label{fig3}
Zoom of the set of zeros in the vicinity of the origin, for 19000 zeros
of the partition function. The density of points corresponds to the
density of zeros. The density of zeros vanishes in the
vicinity of the origin.
\end{figure}

\begin{figure}[h]
\caption{}
\label{fig4}
Magnetization distribution for the hierarchical lattice with 4 generations.
For clarity, the plots have been shifted along the y axis.
$750000$ iterations of Swendsen algorithm were carried out.
\end{figure}

\begin{figure}[h]
\caption{}
\label{fig5}
Construction of the regular simplex.
\end{figure}

\begin{figure}[h]
\caption{}
\label{fig6}
Magnetization distribution for the regular simplex with 7 generations.
For clarity, the plots have been shifted along the y axis.
$750000$ iterations of Swendsen algorithm were carried out.
\end{figure}

\begin{figure}[h]
\caption{}
\label{supp}
Inverse temperature cross-over as a function of the number of generations
for the hierarchical lattice. The cross-over is such that the average
magnetization is some constant, the two extremal spin being frozen in
the up direction. The constant is taken equal to $0.02$.
\end{figure}

\begin{figure}[h]
\caption{}
\label{fig7}
Magnetization distribution for a percolation cluster of size
$N =2660$ sites (this cluster is pictured on figure
\ref{fig8}.The temperature is $T=0.34,0.42,0.50,0.58,0.66,0.74$.
$750000$ iteration
of the Lanczos algorithm were carried out.
\end{figure}

\begin{figure}[h]
\caption{}
\label{fig8}
Two spin configurations in the peak $N_{\uparrow}/N \simeq 0.38$.
\end{figure}

\begin{figure}[h]
\caption{}
\label{fig9}
Magnetization distribution averaged over the geometry. The clusters
are generated in a 60x60 box, and 100000 steps of Swendsen algorithm
are carried out. For $T=0.4 (0.3, 0.5)$,
$106 (200, 58)$ different geometries were generated. Notice the
semi-log scale.
\end{figure}

\begin{figure}[h]
\caption{}
\label{fig10}
Magnetization distribution averaged over the geometry. The clusters
are generated in a 60x60 box and a 25x25 box,
and 100000 steps of Swendsen algorithm are carried out. The temperature
is $T=0.25$.
\end{figure}

\begin{figure}[h]
\caption{}
\label{fig11}
Variations of the correlation $R$ as a function of the temperature
for the cluster of figure \ref{fig8}. The average number of up spins
is chosen to be around $0.38$.
\end{figure}

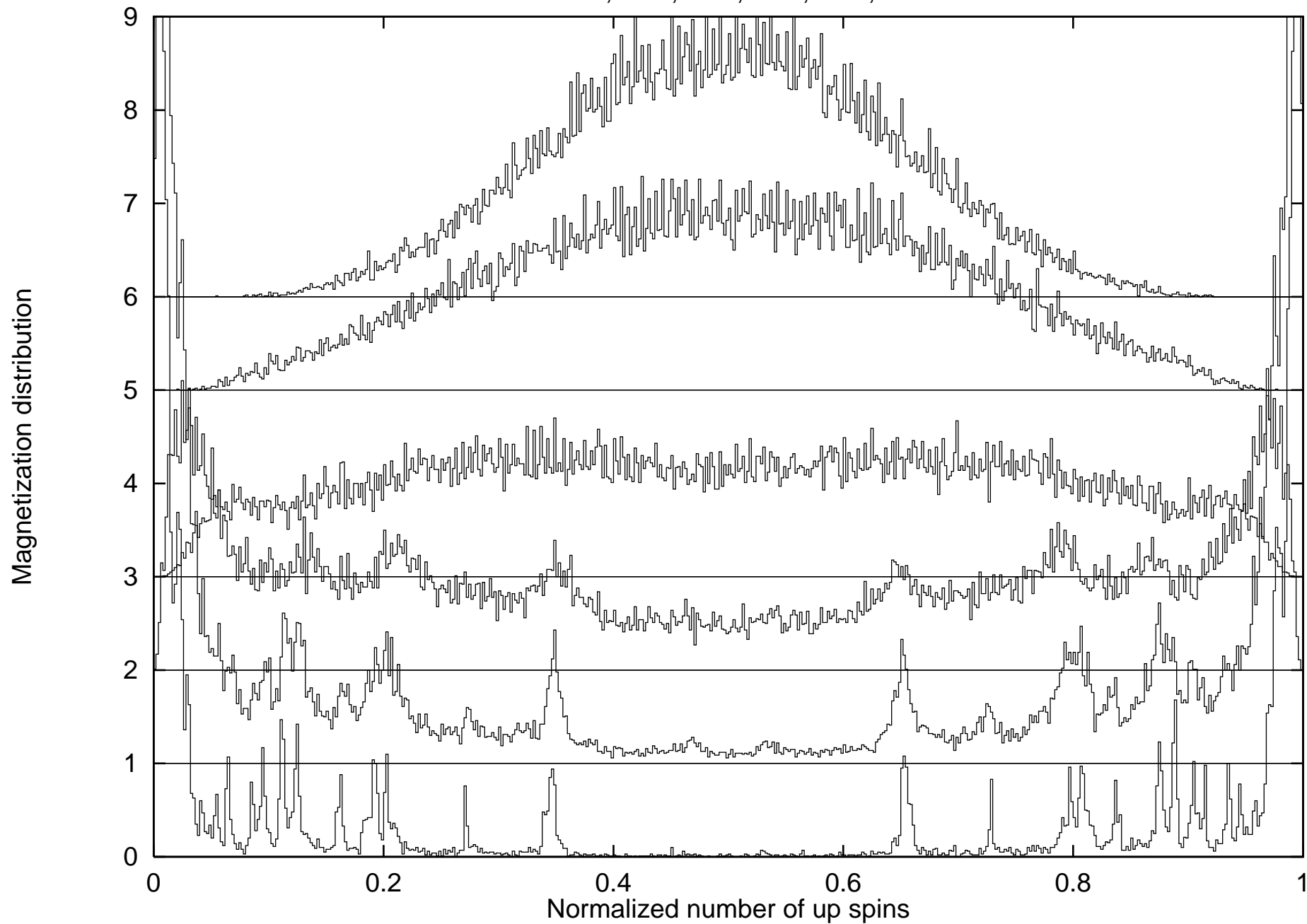
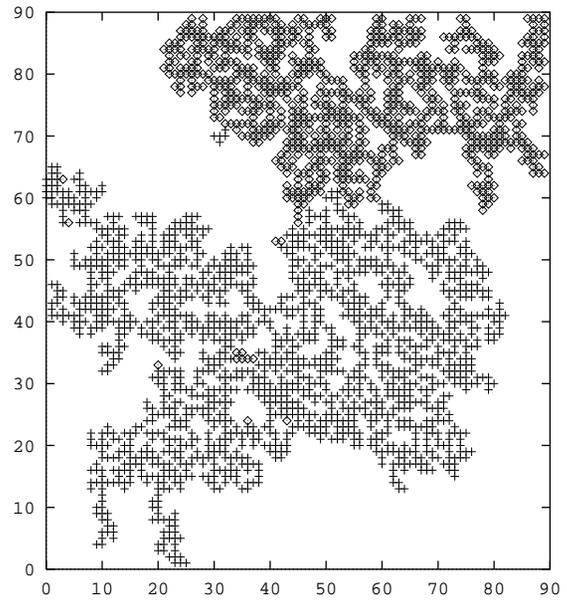
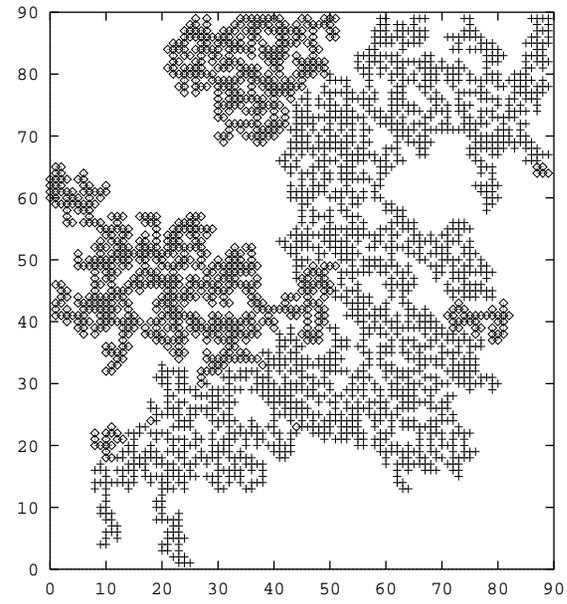
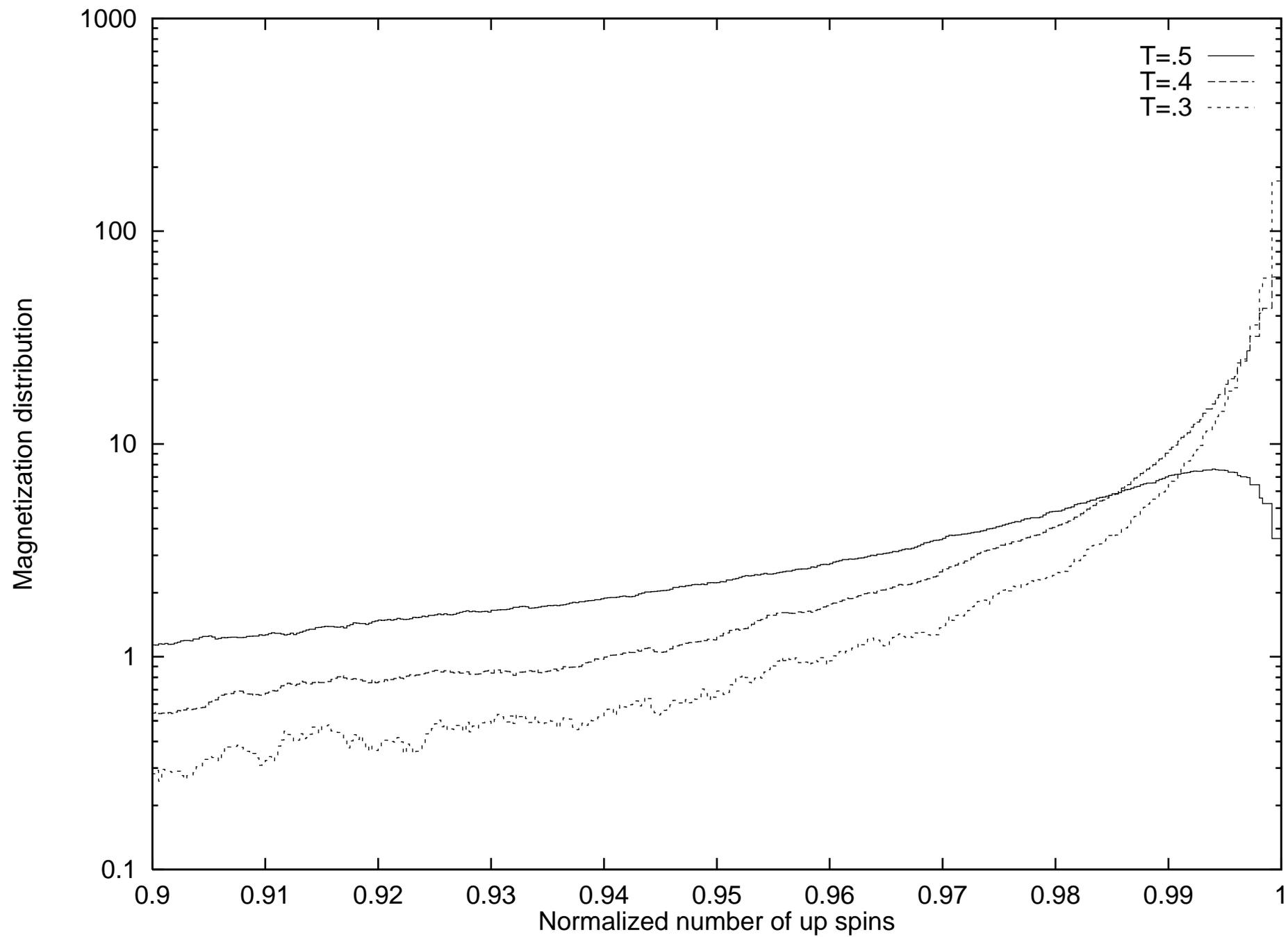
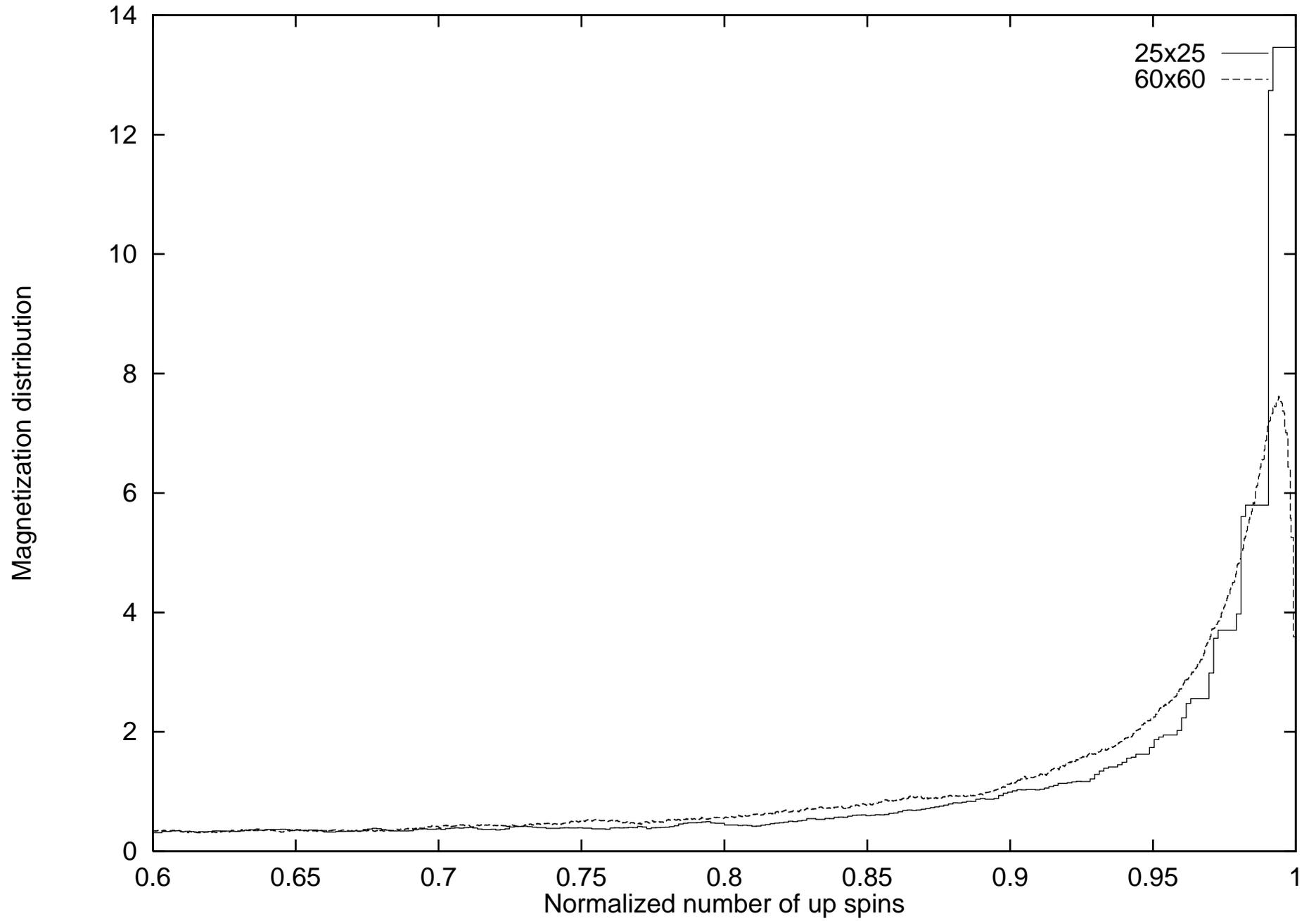
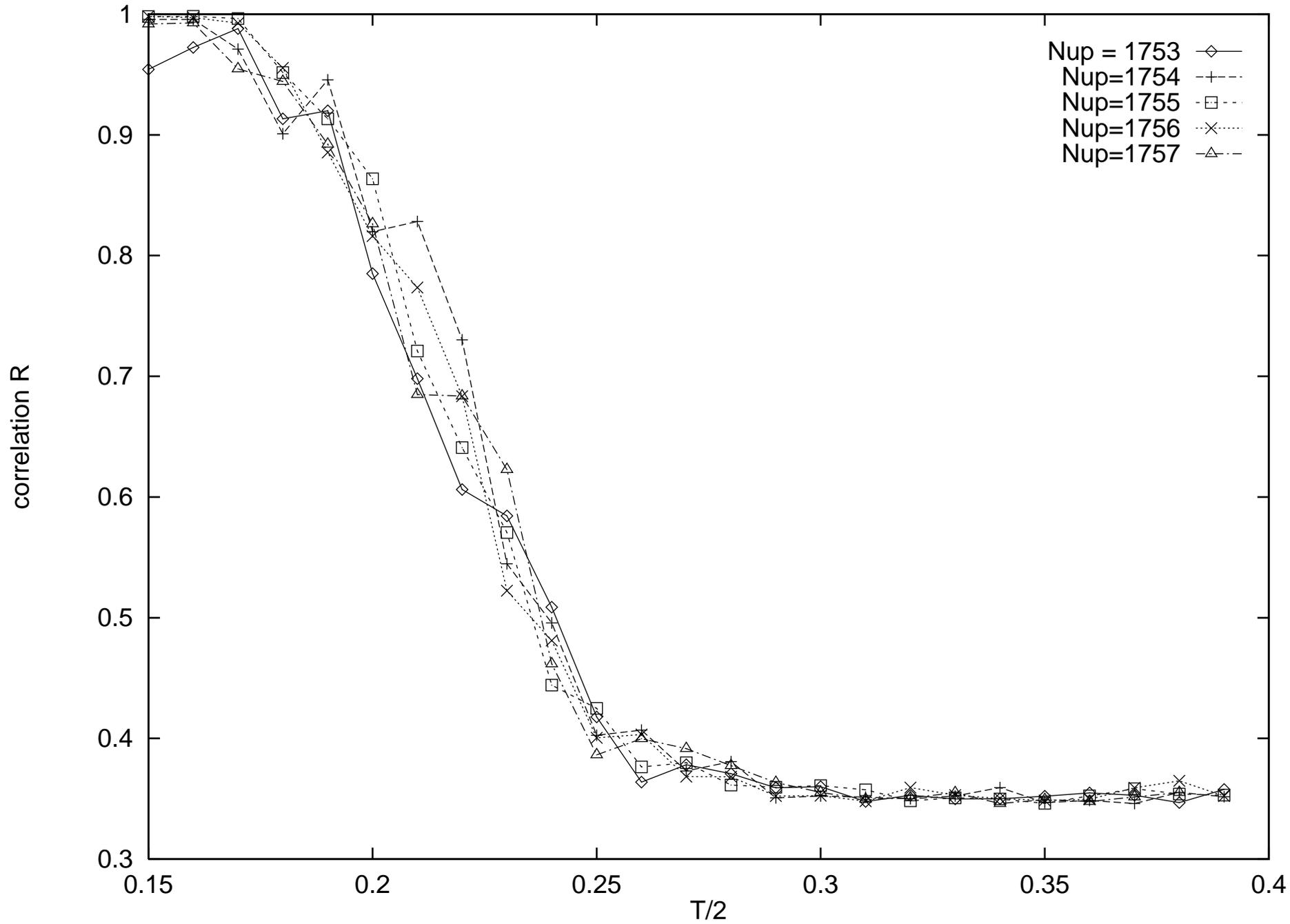
\end{document}